\documentclass[twoside]{article}
\usepackage{fleqn,espcrc2}

\usepackage{graphicx}

\newcommand\photino{\tilde{\gamma}}
\newcommand\gluino[1]{\tilde{g}{}^{#1}_{}}
\newcommand\Charp[1]{\tilde{\chi}{}^{+}_{#1}}
\newcommand\Charpq[1]{\bar{\tilde{\chi}}{}^{+}_{#1}}
\newcommand\Neutr[1]{\tilde{\chi}{}^{0}_{#1}}
\newcommand\Neutrq[1]{\bar{\tilde{\chi}}{}^{0}_{#1}}
\newcommand{\AmS}{{\protect\the\textfont2
  A\kern-.1667em\lower.5ex\hbox{M}\kern-.125emS}}

\title{Light Gauginos -- a Solution to More than the EDMs?}

\author{Thomas Gajdosik\address[UA]{University of Alabama;
        Tuscaloosa, Alabama 35487}%
        \thanks{This work was supported by DOE grant DE-FG02-96ER40967}}

\begin{document}

\begin{abstract}
In this talk I want to present questions that remained unclear to me
in the last years.
These questions concern the Electric Dipole Moments of electron and neutron
and the way people exclude regions of parameter space.
\vspace{1pc}
\end{abstract}

\maketitle

\section{Introduction and Outline}

As this talk was primarily aimed at raising questions that could be
discussed in private after the talk or during the workshop, it is not
suited to be reproduced just as it was given. I will try to incorporate
the discussion and also results from the discussion into this small
article.

In the next section I will discuss light gauginos. I will
present my opinion about the electric dipole moments and their
measurements in the third section. My conclusions will follow in
section four.

\section{Light gauginos}

When supersymmetry (SUSY) was found and
the minimal supersymmetric standard model (MSSM) was introduced
\cite{mssm},
people also looked at possibilities to restrict
the large number of parameters by symmetry arguments. There are two
different symmetry arguments. One argument is based on symmetries
and simple boundary conditions of the renormalization group
equations
\cite{rges}.
The second and more straight forward argument is
based on symmetries in the low energy theory like lepton number
conservation, baryon number conservation, $R$--parity, or a
continuous $R$--symmetry \cite{R-symmetry}.

The continuous $R$-symmetry restricts the soft breaking parameters
quite severely: it puts the trilinear terms and the gaugino mass terms
to zero. This scenario has been discussed a lot:
\cite{light gluino scenarios,Clavelli:1992zv}.
Of course,
with no independent phases left, all SUSY induced CP violation has to
vanish. \cite{Clavelli:2000ua} tried to quantify this vanishing assuming
that the continuous $R$--symmetry is only
an approximate symmetry. But with no continuous $R$--symmetry, what is
the motivation for ''light'' gaugino mass parameters?
% What does ''light'' mean?

\subsection{What does ''light'' mean?}

Performing computations ''light'' means, that expansions in a ratio of
the light mass versus some other mass converges quite quickly. So
''light gauginos'' means actually small gaugino mass parameters. This
does not imply that all the particles are at the same mass as the
parameters. And of course, the gluinos $\gluino{}$ and the lightest
neutralino $\Neutr{1}$ will be of the same order as the corresponding
gaugino mass parameter.

\subsubsection{Charginos}
\begin{figure} \begin{center} \begin{picture}(180,160)(0,0)
%\put(0,-40){\framebox(150,200)}
\put(0,-25){\mbox{\resizebox{!}{180pt}{\includegraphics{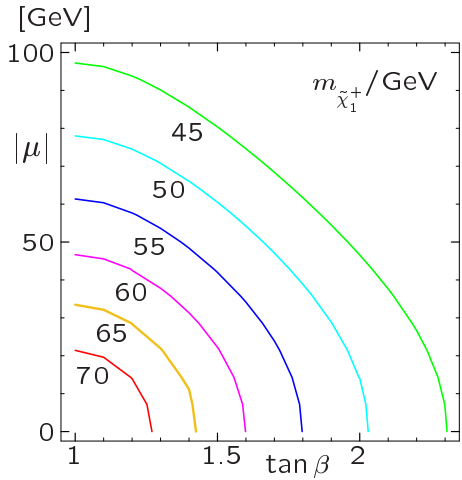}}}}
\end{picture}\\ \end{center}
\caption{%
$\Charp{1}$ mass contours in the $|\mu|$-$\tan\beta\,$ plane for
$m_{SU(2)}=0$.
\vspace{-12pt}
}
\label{fig:mass chargino}
\end{figure}

Let's look for instance at the simple chargino mass matrix:
\begin{equation}
  {\mathcal M}^{\Charp{}}_{}
=
  \left(\begin{array}{cc}
  m_{SU(2)} & \sqrt{2} m_{W}^{} \sin\beta\, \\
  \sqrt{2} m_{W}^{} \cos\beta\, & \mu
  \end{array}\right)
\end{equation}
The term with the $W$--mass sets the scale for the lighter chargino
$\Charp{1}$. Even when $m_{SU(2)} \to 0$, $\Charp{1}$ will be
heavier than $45$~GeV, provided $\tan\beta\,$ will not become too
large, and can nearly become as heavy as the $W$ boson, especially
for small $|\mu|$, see Figure~\ref{fig:mass chargino}.

\subsubsection{Gluinos}
Although there are 8 gluinos, they do not acquire different masses
like charginos or neutralinos. Their mass matrix is just the
$SU(3)$ gaugino mass parameter. So they become massless in the limit
of the continuous $R$--symmetry.

\subsubsection{Neutralinos}
\begin{figure} \begin{center} \begin{picture}(180,160)(0,0)
%\put(0,-40){\framebox(150,120)}
\put(0,-25){\mbox{\resizebox{!}{180pt}{\includegraphics{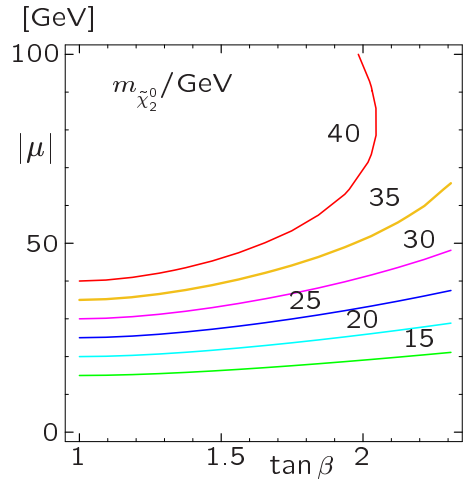}}}}
\end{picture}\\ \end{center}
\caption{%
$\Neutr{2}$ mass contours in the $|\mu|$-$\tan\beta\,$ plane for
$m_{U(1)} = m_{SU(2)} = 0$.
\vspace{-12pt}
}
\label{fig:mass neutralino}
\end{figure}

For the neutralinos the situation is more complicated. When both
gaugino mass parameters that enter the neutralino mass matrix are
exactly zero, an exact photino state $\photino$ with mass zero
decouples from the other neutralinos. The situation remains similar,
if both mass parameters are equal, even if they are not zero. Then
one neutralino will be an exact $\photino$. This $\photino$ does not
couple to the the $Z$ boson at all. So all limits that are derived
from LEP experiments with the analysis of the $Z$ peak do not apply.
But the second lightest neutralino $\Neutr{2}$ becomes then the
particle, that the LEP collaborations were looking for --- aside from
the fact, that it can decay. One can apply now the limits from LEP to
this $\Neutr{2}$. Again $\tan\beta\,$ is quite important. $|\mu|$
works in the opposite direction as with the $\Charp{1}$: now the bigger
$|\mu|$, the heavier $\Neutr{2}$, see Figure~\ref{fig:mass neutralino}.
For a thorough discussion about the neutralino mass matrix see
\cite{Bartl:1989ms}.

\subsubsection{1--loop mass corrections}
It is well known, that the mass corrections to the higgs boson are
quite sizeable. As similar graphs are participating, one can expect
similar corrections for the gauginos. But there is one big difference
to the higgs corrections, where one calculates a $\delta m^{2}_{h}$
that is proportional to $m_{t}^{4}$, whereas $\delta m^{}_{\lambda}$
is only proportional to $m_{t}^{3}$, so that the corrections are not up
to 50~GeV, but only up to 1 or 2~GeV.

One can also expect that the corrections to the $\gluino{}$ are bigger
than to the $\photino$ because of the strong coupling. This is the
reason, why one expects the $\photino$ to be the lightest
supersymmetric particle (LSP) in this scenario.

\subsection{Why are they not seen yet?}
This question is more tough, but there are answers and arguments, that
can be brought up. The first point is, that these light gauginos really
escape detection --- just like the neutrinos escaped detection for a long
time. I will be more specific for each of the gauginos later. The second
point is, that they can also be misinterpreted as other particles. This
argument is used by many people for various other effects, too
\cite{c+t,disguise}.
And there are measurements, that would favor the
existence of light gauginos
\cite{Clavelli:1992zv,c+t,light gauginos,Clavelli:1999sm,Clavelli:1993xq},
see especially the discussion in \cite{Clavelli:1999sm}.

\subsubsection{Charginos}
As $\Charp{1}$ is lighter than the $W$-boson, it must have been produced
by LEP. Looking at the cross section $e^{+} e^{-} \to X$, one expects to
see a threshold when $\Charp{1} \Charpq{1}$ is pair produced. But since
there is no fine spaced energy scan done by LEP above the $Z$-peak, one
can easily miss the threshold. Of course, there has to be a higher
cross section for $e^{+} e^{-} \to X$ above the $\Charp{1} \Charpq{1}$
threshold. But how do the LEP experiments calibrate their detectors, if
not by that cross section, which is assumed to be the standard model one?
The comparison to the Monte Carlo Data is done by the same procedure, too.
My conclusion: it is possible, that the pair production of
$\Charp{1} \Charpq{1}$ is not clearly visible just by looking at the
production cross section.

Ok, so lets look into the decays! At least they should be seen in LEP,
right? If I take the scenario of tiny gaugino mass parameters for real,
I will have to look at the signature of $\Charp{1}$ decay, because it
can be dramatically different from the normal decay. \cite{Clavelli:susy2k}
proposes the hadronic decay $\Charp{1} \to u \bar{d} \gluino{}$ to be
dominant.
That decay would then just increase the hadronic cross section. To see
this increase in the hadronic cross section, one would have to make
either a fine spaced energy scan, to see the threshold or to know exactly
the expected standard model background. Of course, the standard model
background is quite well known.

But what is the procedure, with which the hadronic cross section is
compared to the theoretical prediction? It is the Monte Carlo studies
for the detectors; and they compare to the measured cross section, making
the assumption, that it is the standard model cross section. In some sense
this seems to me like a vicious circle. But I also think these complicated
procedures are necessary, to extract reliable data from the complicated
detectors. And the experimentalists are doing a marvellous job!

\subsubsection{Gluinos}
The $\gluino{}$ is expected to be very light. As such a light colored
state it will affect the running of $\alpha_{s}$. This change of running
is actually in quite good agreement with the measurements performed with
$q \bar{q}$ bound states
\cite{Clavelli:1993xq,Clavelli:1995mf},
but it is off for the measurement done at the $\tau$
\cite{Buskulic:1993sv}.
Nevertheless, assuming a 10\% relativistic correction for the
$\tau$-measurement brings this measurement in the correct range for
the light gluino prediction. But it needs a 100\% correction to bring the
$q \bar{q}$ bound state measurements into
the correct range of the standard model prediction. Of course, both
predictions assume the measurement at the $Z$-peak to be correct.

The Aleph collaboration at LEP did an additional study for the running
of $\alpha_{s}$ and the number of light flavors \cite{Barate:1997ha}. It
ruled out a light $\gluino{}$, but this study was criticized by
\cite{farrar}. Although I don't understand
\cite{Barate:1997ha} fully,
especially since they rely heavily on the Monte Carlo studies and
sophisticated statistical techniques, I think they did a good work.
But one question remains for me: why did the other LEP collaborations
neither confirm nor contradict this analysis?

\subsubsection{Neutralinos}
Here we have to look at two of them: the $\photino$, which is assumed
to be the LSP, and the $\Neutr{2}$.

I thought I could ignore the question about the $\photino$ because it
will not couple to the $Z$ boson. Actually it will couple exactly the
same like an ordinary photon, with the only difference, that one of
the other particles has to be a SUSY particle.
Since SUSY particles are quite heavy, i.e.
much heavier than the normal quarks and leptons, I assumed, that the
effects would be too small to be seen anyway. But during the workshop
I was made aware that there exist dedicated searches for a light
$\photino$ in the low energy $e^{+} e^{-}$ cross section
\cite{photino search 1}.
But these experimental results do not
exclude a light $\photino$, they just give limits on the mass
of the scalar electrons. Other restrictions for a $\photino$ can be
found from cosmological arguments \cite{photino search 2} and
other inclusive detector measurements
\cite{photino search 3}.

For the $\Neutr{2}$ the situation is quite complicated. One has to worry
not only about the production of $\Neutr{2}\Neutrq{2}$ at the $Z$ peak,
one has also to think about possible decays. Again \cite{Clavelli:susy2k}
proposes decay modes into $\gluino{}$ and hadrons. These decay modes
look similar to the decays of the $\Neutr{1}$ in $R$-Parity violating
models \cite{RPV,Bartl:2000yh}. Keep in mind, that LEP did extensive
studies for $R$-Parity violation! But on the other hand the situation
is not the same: there is no simple way to translate limits obtained
for $R$-Parity violating models into limits for this hadronic
$\Neutr{2}$ decay, since one of the main features of the $R$-Parity
violating models can be the enhancement in the lepton multiplicity
\cite{Bartl:2000yh}. But
the hadronic decay of $\Neutr{2}$ will reduce the lepton multiplicity.

\subsubsection{My opinion on light gauginos}
Whenever I asked experimentalists about excluding the light gaugino
scenario by experimental data, I got the answer, ''We would have to
make a dedicated study''. Only G.~Dissertori gave a definite answer.
So I --- for myself --- cannot rule out that
light gaugino scenario. And in addition, there are some problems,
where light gauginos offer a solution.

\subsection{Motivation for light gauginos}
When one looks at the unconstrained MSSM that has all soft breaking
parameters with arbitrary phases, one will immediately have problems
with existing measurements on $CP$ violating variables and on
flavor changing processes. The usual way to handle this kind of
situation is to apply constraints on the parameters, either by
using renormalization group equations with restrictive boundary
conditions or applying low energy symmetry arguments like ''alignment''
\cite{Alignement,Gerard:1984bg}.

On the other hand, light gauginos are another way of suppressing many
of the problematic diagrams that give too big results for $CP$ violating
observables. \cite{Clavelli:2000ua} has quantified this argument for the
electric dipole moments (EDMs) of electron and neutron. In
\cite{Gerard:1984bg} one can
see the vanishing of the SUSY effects in $K^{0}$-$\bar{K}^{0}$ mixing
with vanishing gaugino masses. \cite{irina} has demonstrated this
vanishing also for a top quark mass of 175~GeV.

One motivation for me to look for light gauginos is, that almost
everybody seems inclined to consider the relevant parameter space to
be already ruled out, but no one could really convince me about that.

\section{Electric dipole moments}
The second big question I wanted to present in the workshop was about
EDM measurements. It can be brought to the point, that I have doubts
about the prerogatives that are assumed when a macroscopic measurement
is compared to calculations that involve questions of low energy QCD.
My feeling was --- and is to some extent even now --- that the accuracy
of the QCD related topics is not the same as the claimed
accuracy of the experiments that measure the EDMs: do we really
know atomic or nuclear physics to the precision of $10^{-12}$, which
is the inherent precision of the EDMs. A discussion about that point
with Maxim Pospelov during the workshop was very helpful for me.

\subsection{EDM of the neutron}
The basic question about the calculation of the EDM of the neutron
$d^{}_{N}$ is, how one can combine the result of the various
contributions of the particles, that make up the neutron. The first
estimates were based on the non-relativistic SU(6) quark model with
\begin{equation}
d^{}_{N} = (4/3) d^{d}_{} - (1/3) d^{u}_{}
\enspace .
\end{equation}
And the other QCD contributions to $d^{}_{N}$ were estimated by a
Naive Dimensional Analysis \cite{Manohar:1984md}. Different estimates were
made in \cite{Ellis:1996dg} and a few months ago I found even another
proposal \cite{DQQM}. This multitude of possible models, that give
opposite results for $d^{}_{N}$ --- as has been shown in \cite{Bartl:1999bc}
--- shows that the theoretical part of the measurements is not
really under control. So any restriction derived from the
non measurement of $d^{}_{N}$ can just be applied if one makes
additional assumptions that are not part of the MSSM.

In the workshop I was made aware of more accurate calculations for
the composition of the neutron
\cite{NEDM-QCD},
see the discussion
in \cite{Falk:1999tm}. A new estimate of the chromoelectric contribution
has been obtained in \cite{Pospelov:2000bw}.
So some part of my criticism of the neutron EDM has
lost its basis. But the possibility of cancellations is still present.
And the points in parameter space where these cancellations take place
depend on the specific assumptions about the neutron.

\subsection{EDM of the electron}
In contrast to $d^{}_{N}$ the EDM of the electron $d^{}_{e}$ is easy
to calculate and theoretically fully under control. The problematic
point comes with the measurement, because a direct measurement with free
electrons seems to be impossible. One argument for the impossibility is
the charge of the electron.

The usual way is to measure atomic quantities and to relate them to
$d^{}_{e}$. The most accurate measurements today are done with
$d^{}_{a} ( {}^{205}\mathrm{Tl}; 6 \, {}^2 P_{1/2} )$ \cite{eedm}.
The relevant calculations were done in \cite{Barr:1992cm}, pointing out,
that $d^{}_{\mathrm{Tl}}$ has four sources and only one of them is
the intrinsic $d^{}_{e}$. The other three are
(1) an intrinsic nucleon EDM,
(2) $P,T$-odd nucleon-nucleon interaction,
(3) $P,T$-odd electron-nucleon interaction.
Non measurement of $d^{}_{\mathrm{Tl}}$ requires the sum of all
four sources to be smaller than the experimental limit.

To derive a limit on $d^{}_{e}$ from the $d^{}_{\mathrm{Tl}}$
measurement one has to assume, that all other three contributions
are much smaller then the first. Of course, the probability that
two of these four contributions cancel each other is very small.
But on the other hand: more contributions mean more
possibilities for cancellations.

\subsection{EDM of ${}^{199}\mathrm{Hg}$}
I was made aware of this type of measurement, that restricts
the parameters of the MSSM, by Toby Falk during SUSY99 and then
later again by Maxim Pospelov at this workshop, see their article
\cite{Falk:1999tm}.

As it uses again
atomic and nuclear physics to relate the measurement to the EDMs of
the constituents of the atom, my uneasy feeling about the accuracy
remains. But it is another argument, that has to be taken into account.
Any model should describe all phenomena that are encountered in
nature, so it should explain the non-measurement of
$d_{\mathrm{Hg}}^{}$, too.

\section{My conclusions}
{\em ''\dots the supersymmetry algebra is the only graded Lie algebra of
symmetries of the $S$-matrix consistent with relativistic quantum
field theory''} \cite{bagger & wess}. This citation reflects the
major motivation for SUSY. Since we live in a world where masses
are nonzero, it is necessary that SUSY is broken. But this breaking of
SUSY introduces a lot of new parameters in the low energy Lagrangian,
that are constrained by stability requirements and their effect on
measurements.

Stability requirements just restrict the parameters in a way, that the
vacuum of the model is compatible with what we see in everyday life:
we don't see any electric or colored net charge in the vacuum. And we
see that particles are massive, but not superheavy (of order of the
Planck mass). And we see, that the electro-weak symmetry is broken
to the electric charge. So the vacuum of SUSY has to fulfill these
constraints, which it does easily enough with the mechanism of
spontaneous symmetry breaking in the electro-weak sector.

But there are a lot of other effects, that wait for an explanation:
we see a large baryon asymmetry in the universe (BAU), but the vacuum of
the Lagrangian of the MSSM looks symmetric between baryons and
antibaryons. And the proton is quite stable, too. And we also find,
that $CP$ violation is quite small. Actually, the only visible effect
up to now is observed in the neutral kaon sector \cite{cp-experiments}.

For the explanation of BAU one needs quite a large $CP$ violating phase
\cite{baryogenesis},
but other measurements, like the EDMs, can be
interpreted as restricting these possible phases --- or at least specific
combinations of them. Another usual interpretation is, that the masses
of the SUSY particles are so heavy, that their effect in the EDMs are
suppressed. But having very heavy SUSY particles makes the explanation
of the observed vacuum more difficult.

Another interpretation for the non-observation of EDMs can be
light gauginos \cite{Clavelli:2000ua}. It also allows the other SUSY
particles to stay ''light'' and helps therefore in the explanation of
the electro-weak symmetry breaking. The ''cost'' for this
interpretation is,
that one is away from the mainstream of model building. And one has
to investigate carefully, if other measurements will not rule out
this scenario.

\end{document}